\documentclass[11pt,showpacs,aps]{revtex4}
\usepackage{dcolumn}
\usepackage{bm}
\usepackage{amsmath}

\begin{document}
\title{Instability of the Perturbation Theoretical Chromodynamic Vacuum}

\author{Y.N. Srivastava, O. Panella}
\affiliation{Physics Department \& INFN, University of Perugia, Via
Pascoli, Perugia Italy}
\author{A. Widom}
\affiliation{Physics Department, Northeastern University, Boston MA, USA}

\begin{abstract}
The standard model of strong interactions invokes the
quantum chromodynamics (QCD) of quarks and gluons interacting within
a fluid. At sufficiently small length scales, the effective
interactions between the color charged particles within the fluid
are thought to be weak. Short distance asymptotic freedom provides
the perturbation theory basis for comparisons between QCD theory and
laboratory high energy scattering experiments. It is here shown that the
asymptotically free vacuum has negative dissipation implicit in the
color electrical conductivity. Negative dissipation implies an
asymptotically free QCD negative temperature {\em excited state amplifier}
unstable to decay. The qualitative experimental implications of this
instability are explored.
\end{abstract}


\maketitle

The physical vacuum~\cite{Wilson:1974} in both quantum
electrodynamics (QED) and quantum chromodynamics (QCD) is endowed
with a radiation impedance, in Lorentzian units \begin{math}
R_{vac}\equiv 1/c \end{math}, an electric charge
\begin{math} e \end{math} and a color charge \begin{math} g \end{math}. The
weak coupling of strength of QED and the strong coupling strength of
QCD are described by the vacuum properties
\begin{eqnarray}
\alpha &=& \left( \frac{R_{vac} e^2}{4\pi \hbar }\right)
=\frac{e^2}{4\pi\hbar c}\ ,
\nonumber \\
{\alpha}_s &=& \left( \frac{R_{vac} g^2}{4\pi \hbar }\right)
=\frac{g^2}{4\pi\hbar c}\ .
\label{coupling}
\end{eqnarray}
The quantum vacuum of both QED and QCD is far from being
inert~\cite{Gribov:1999,Srivastava:2001,Srivastava:2001b,Grau:2004}
when probed at microscopic distances and times. For the relativistic
wave vector \begin{math} q=(\omega/c, \bm{k}) \end{math}, one may
speak of space-like vacuum correlation at an invariant wave number
\begin{math} Q\equiv \sqrt{-q^2}>0 \end{math} wherein
\begin{equation}
Q^2= -q^2 =  |\bm{k}|^2 -({\omega^2}/{c^2}).
\label{space_like}
\end{equation}
One may also speak of time-like vacuum
dynamics~\cite{Shirkov:1997,Nesterenko:2000} at an invariant
frequency
\begin{math} \Omega \equiv c\sqrt{s}>0 \end{math} wherein
\begin{equation}
s= q^2 = (\omega^2/c^2) - |\bm{k}|^2.
\label{time_like}
\end{equation}

The complete vacuum dielectric \begin{math} \varepsilon (Q^2)  \end{math},
respectively color dielectric \begin{math} \varepsilon_s (Q^2)  \end{math},
response function and the complete magnetic permeability
\begin{math} \mu (Q^2)  \end{math}, respectively color magnetic permeability
\begin{math} \mu_s (Q^2)  \end{math},
describe a dynamic vacuum exhibiting charged, respectively color
charged, fluctuating quantum currents. Both QED and QCD theory
envision~\cite{Berestetskii:1982} a dynamic vacuum in terms of a
{\it Q}-dependent coupling strength for space-like correlations
\begin{eqnarray}
\alpha(Q^2) &=& \frac{e^2}{4\pi \hbar c\varepsilon(Q^2)}
=\frac{e^2\mu (Q^2)}{4\pi \hbar c}\ ,
\nonumber \\
\alpha_s(Q^2) &=& \frac{g^2}{4\pi \hbar c\varepsilon_s(Q^2)}
=\frac{g^2\mu_s (Q^2)}{4\pi \hbar c}\ .
\label{CouplingStrength}
\end{eqnarray}
Crucially important for what follows is that QED,
respectively QCD, may also describe vacuum time-like
dynamical fluctuations by employing an electrical conductivity
\begin{math} \sigma  \end{math}, respectively a color
electrical conductivity \begin{math} \sigma_s  \end{math}.
\begin{eqnarray}
\varepsilon (s+i0^+) &=& 1+\frac{i\sigma (s)}{c\sqrt{s}}\ ,
\nonumber \\
\varepsilon_s (s+i0^+) &=& \frac{i\sigma_s (s)}{c\sqrt{s}}\ .
\label{conductivity}
\end{eqnarray}
Causality dictates dispersion relations between the
dielectric response functions and the dynamic conductivities;
Specifically
\begin{eqnarray}
\varepsilon (Q^2) &=& 1-\left(\frac{Q^2}{\pi c}\right)
\int_0^\infty  \frac{\Re{e}\{\sigma(s)\}\ ds}{s(s+Q^2)\sqrt{s}}\ ,
\nonumber \\
\varepsilon_s (Q^2) &=& -\left(\frac{Q^2}{\pi c}\right)
\int_0^\infty  \frac{\Re{e}\{\sigma_s(s)\}\ ds}{s(s+Q^2)\sqrt{s}}\ .
\label{dispersion}
\end{eqnarray}
The central theorem of this work involves the nature of
\begin{math} \alpha(Q^2) \end{math} and the
\begin{math} \alpha_s(Q^2) \end{math} when analytically continued from
the space-like \begin{math} Q^2>0 \end{math} regime to the time-like
\begin{math} s>0 \end{math} regime wherein the complex coupling strengths
are dynamic.
\smallskip \par \noindent
\textbf{Theorem:} \emph{For time-like wave vectors \begin{math} s>0 \end{math},
the sign of the real part of the conductivity is opposite to the sign of the
imaginary part of the coupling strength.}
\smallskip \par \noindent
\textbf{Proof:} Eqs.(\ref{CouplingStrength}) and (\ref{conductivity}) imply
\begin{eqnarray}
-{\Im m}\{\alpha(s+i0^+)\} &=&
\left|\frac{\alpha(s+i0^+)}{\varepsilon (s+i0^+)}\right|
\frac{\Re{e}\{\sigma(s)\}}{c\sqrt{s}}\ ,
\nonumber \\
-{\Im m}\{\alpha_s(s+i0^+)\} &=&
\left|\frac{\alpha_s(s+i0^+)}{\varepsilon_s (s+i0^+)}\right|
\frac{\Re{e}\{\sigma_s(s)\}}{c\sqrt{s}}\ .
\label{ProofTheorem}
\end{eqnarray}
from which the theorem follows.
\smallskip \par \noindent
(i) In standard QED perturbation theory,  the
vacuum conductivity obeys a condition of positive dissipation
\begin{equation}
{\Im m}\{\alpha(s+i0^+)\}\le 0 \ \ \ \Rightarrow
\ \ \ \Re{e}\{\sigma(s)\}\ge 0.
\label{QEDdissipate}
\end{equation}
The \emph{{vacuum heating}} induced by an applied electromagnetic
field at high frequency is due to the creation of one or more pairs
of oppositely charged particles and anti-particles.
\par \noindent
(ii)
In standard QCD perturbation theory, the vacuum color conductivity
obeys a condition of negative dissipation, i.e. positive amplification
\begin{equation}
{\Im m}\{\alpha_s(s+i0^+)\}\ge 0 \ \ \ \Rightarrow
\ \ \ \Re{e}\{\sigma_s(s)\}\le 0.
\label{QCDamplify}
\end{equation}
Our purpose is to point out that negative dissipation in the form
\begin{math} \Re{e}\{\sigma_s (s)\} < 0 \end{math}
is implicit in {\em all} previously reported QCD perturbation
analyses to finite order in \begin{math} \alpha_s \end{math}.

The physical interpretation of QCD {\em vacuum cooling}
induced by the application of a high frequency
chromodynamic field is (at first glance) more than a little
bit obscure. How might one cool what is already asserted to
be the vacuum? A true quantum vacuum is conventionally
visualized as having the lowest possible energy. No further
cooling should then be possible. Asymptotic
freedom~\cite{Politzer:1973,Gross:1973,Wilczek:1999} must
refer to an excited state color amplifier. One may indeed
spontaneously cool down an initially  excited state. We
have been forced to conclude that the asymptotically free
QCD vacuum is in reality unstable to decay into the true
vacuum state. However, the situation with regard to
comparisons between theory and experiment may not be
entirely desperate.

It is important to note that negative dissipation implicit in
condensed matter electrical conductivity is by no means a new
phenomenon in laboratory amplifying systems~\cite{Panella:2006}.
If some of the chemical
substances within a condensed matter system have atoms or molecules
in {\em inverted} populations of excited state electronic energy
levels, then the electrical conductivity of the substance at hand
(within an amplifier frequency bandwidth \begin{math} \varpi \end{math})
can exhibit a negative dissipative part of the electrical conductivity,
i.e.
\begin{math}
\Re{e}\{\sigma_{Amplifier} (\omega \in \varpi) \} \leq 0
\end{math}.
Such systems\cite{Ramsey:1956,Klein:1956} exhibit a negative noise
temperature
\begin{math}
T_{n}(\omega \in \varpi) < 0
\end{math}
within the same bandwidth \begin{math} \varpi \end{math}. The
laboratory achievement of negative noise temperatures,
characteristic of electronic energy levels with inverted
populations, was absolutely essential in order to achieve the
original maser and laser
devices~\cite{Schalow:1958,Weber:1959,Maiman:1960}. Negative noise
temperatures also occur naturally in the astrophysical matter
clouding
galaxies~\cite{Cohen:1989,Herman:1985,Eliztur:1982,Weaver:1965}. The
negative dissipative part of the electrical conductivity in some
astrophysical matter is made manifest via the measured negative
absorption (i.e. amplification) of the electromagnetic radiation
that passes through these materials. However, it must be realized
that amplification due to excited state systems has only a finite
lifetime. The excited state atoms decay into their ground states
after giving rise to a burst of electromagnetic radiation. To again
achieve the excited state amplifier status, energy must be pumped
back into the matter by an external source.

We note that, Savvidy and Matinyan  showed, a long time ago, that a constant color magnetic field is spontaneously generated in QCD~\cite{Savvidy:1977as,Matinyan:1976mp,Batalin:1976uv,Branchina:1992nx}.

They interpreted it to imply an infrared instability
inherent in QCD. On the other hand, the instability
discussed in the present paper arises directly from the
asymptotic freedom expression for the QCD coupling constant
and is apparently quite different.

For the QCD problem at hand, let \begin{math} \Lambda \end{math}
denote the wave number beyond which asymptotic freedom might be
thought to prevail. Perturbation theory
predictions~\cite{Politzer:1973,Gross:1973} for the QCD coupling
strength appear in the form
\begin{equation}
\frac{1}{\alpha_s(Q^2)} = \left[\frac{33-2N_f}{12\pi}\right]
\ln \left[\frac{Q^2}{\Lambda^2}\right],
\label{perturbation}
\end{equation}
wherein \begin{math} N_f=5 \end{math} is the present experimental
number of {\em light} quark flavors. The QCD question raised by us
concerns the negative dissipative part of the color conductivity.
We find from Eqs.(\ref{CouplingStrength}), (\ref{ProofTheorem}) and
(\ref{perturbation}) that
\begin{math}
{\Re e}\left\{\sigma_s (s)\right\} < 0
\end{math}
for \begin{math} s > \Lambda^2 \end{math}.
The negative dissipation is implicit in \textit{all} previous
theoretical asymptotic freedom perturbation theory estimates. For
both QED and QCD, the causal dispersion Eq.(\ref{dispersion}) yields
conceptual problems for quantum field theory.

The vacuum screened Coulomb law potential energy between two charges,
\begin{math} Z_1e \end{math} and \begin{math} Z_2e \end{math},
may be written
\begin{eqnarray}
U(r) &=& \frac{Z_1Z_2e^2}{4\pi r}\chi(r),
\nonumber \\
\chi(r) &=& \frac{2}{\pi }\int_0^\infty
\frac{\sin(Qr)}{Q\,\varepsilon(Q^2)} dQ.
\label{CoulombPotential}
\end{eqnarray}
To achieve the Coulomb law of force between two point electrical charges
one requires that
\begin{equation}
\lim_{r\to \infty} \chi(r)=\lim_{Q^2\to 0}\ \frac{1}{\varepsilon(Q^2)}=1.
\label{ScreenQED}
\end{equation}
consistent with the dispersion Eq.(\ref{dispersion}) for
\begin{math} \varepsilon (Q^2)  \end{math}. While the QED perturbation
theory conductivity obeys the expected positive dissipation condition
\begin{math} {\Re e}\{ \sigma(s)\}\ge 0 \end{math}
in the time-like region \begin{math} s>0 \end{math}, the dispersion
relation yields a change in sign for
\begin{math} \varepsilon (Q^2)\end{math} in the space-like region
\begin{math} Q^2>0 \end{math}. There exists a positive wave number
\begin{math} K > 0 \end{math} for which
\begin{math} \varepsilon (K^2)=0\end{math}; i.e. there appears a
Landau ghost \cite{Landau:1955,Sivasubramanian:2002}. For space-like
wave vectors \begin{math} Q^2>K^2 \end{math}, two particles whose
electric charges have the same sign will tend to attract rather than
repel one another. If within even a limited range of wave numbers
attraction occurs between particles whose charges have the same
sign, then there will be a strong tendency towards Cooper
pairing~\cite{Rajagopal:2001} and superconductivity. The
experimental QED vacuum is an insulator rather than a
superconductor. One tends to regard the Landau ghost as an
unphysical~\cite{Bogoliubov:1959} theoretical QED embarrassment.

The vacuum confining potential energy between two color
{\em matrix charged} particles,
\begin{math} g{\sf T}_1 \end{math} and
\begin{math} g{\sf T}_2 \end{math}, may be written\cite{Parihar:2006}
\begin{eqnarray}
U_s(r) &=& \frac{\eta^{ab} T_{a1} T_{b2} g^2}{4\pi r}\chi_s(r),
\nonumber \\
\chi_s(r) &=& \frac{2}{\pi }\int_0^\infty
\frac{\sin(Qr)}{Q\, \varepsilon_s(Q^2)} dQ.
\label{ColorPotential1}
\end{eqnarray}
By taking two derivatives of Eq.(\ref{ColorPotential1}),
\begin{equation}
\chi_s^{\prime \prime }(r)=-\frac{2}{\pi }\int_0^\infty
\frac{Q\,\sin(Qr)}{\varepsilon_s(Q^2)} dQ,
\label{ColorPotential2}
\end{equation}
one finds the finite limits
\begin{eqnarray}
\frac{L^2}{2} &=& \lim_{Q^2\to 0} \frac{\varepsilon_s(Q^2)}{Q^2}
=-\left(\frac{1}{\pi c}\right)
\int_0^\infty  \frac{\Re{e}\{\sigma_s(s)\}}{s^{5/2}}ds\ ,
\nonumber \\
\frac{2}{L^2} &=& -\lim_{r\to \infty}\chi_s^{\prime \prime }(r)
=-\lim_{r\to \infty} \frac{2\chi_s(r)}{r^2}\ ,
\label{ColorPotential3}
\end{eqnarray}
in agreement with the color dielectric function dispersion
Eq.(\ref{dispersion}). The central implication of Eqs.(\ref{ColorPotential1})
and (\ref{ColorPotential3}) is the linear confining potential
\begin{equation}
U_s(r)=-(\eta^{ab} T_{a1} T_{b2})\tau r
\ \ \ {\rm as}\ \ \ r\to \infty.
\label{ColorPotential4}
\end{equation}
Eq.(\ref{ColorPotential4}) may be viewed as a potential due to
a QCD color electric flux tube, i.e. a QCD {\em string} with tension
\begin{equation}
\tau = \frac{g^2}{4\pi L^2}
\label{QCDstring}
\end{equation}
connecting two color charged particles. The above argument is
similar to that given by Feynman~\cite{Feynman:1974} who conjectured
that the color confining potential should obey the bi-harmonic
equation. The fact that the Landau ghost does not haunt QCD
perturbation theory~\cite{Lee:1981} is {\em totally} dependent upon
the negative dissipation in the color conductivity. The resulting
asymptotic freedom QCD vacuum instability might be equally
embarrassing as the Landau ghost. But perhaps a QCD amplifier state
is not unphysical.

Physical negative dissipation can occur when matter is pumped up by
an external energy source. The resulting amplifier excited state can
then return the energy when the matter decays back into the true
ground state. In laboratory scattering experiments with large
incident energies, an initial collision can excite the quark gluon
constituents~\cite{Feynman:1969,Bjorken:1969} as well as pump up a
small space-time piece of the vacuum. Within that small space-time
collision region the QCD perturbation theory may have some validity
while the vacuum is pumped into an excited state. However, the
outgoing fragments of a collision must then contain energy in part
due to the decay of the asymptotic free vacuum back to the true
vacuum. The QCD view of strong interactions might then be flawed
only to the extent that one has no clear picture of the true vacuum
present at times well after the out going fragments (hadrons) enter
into laboratory detectors.

\end{document}